\documentclass[manuscript, nonacm]{acmart}

\usepackage{algorithm}
\usepackage{algorithmic}
\usepackage{multirow}
\usepackage{multicol}
\usepackage{arydshln}
\usepackage{multirow}
\usepackage{longtable}
\usepackage{array}
\usepackage{colortbl}
\usepackage{xcolor}
\usepackage{soul}
\usepackage{changepage}
\usepackage{enumitem}
\usepackage{ragged2e}

\usepackage{tikz}

\definecolor{mediumgray}{HTML}{808080}

\definecolor{orange}{HTML}{FFDFB5} 
\definecolor{green}{HTML}{D2E9B3}
\definecolor{yellow}{HTML}{F7F8A7}
\definecolor{purple}{HTML}{C0A8D9}
\definecolor{pink}{HTML}{EF9DB5}
\definecolor{blue}{HTML}{A1CBFB}
\definecolor{VA}{HTML}{FFDFB5} 
\definecolor{design}{HTML}{F7F8A7}
\sethlcolor{design} 
\newcolumntype{L}{>{\raggedright\arraybackslash}p{2.7cm}}

\definecolor{custompurple}{HTML}{a275ab}
\definecolor{custompink}{HTML}{f27d9b}

\definecolor{lightgray}{gray}{.5}

\newcommand{\lighthline}{\arrayrulecolor{lightgray}\hline\arrayrulecolor{black}}

\newcolumntype{M}{>{\ttfamily\small}p{14cm}}
\newcolumntype{S}{>{\raggedright\arraybackslash}p{5cm}}
\newcolumntype{L}{>{\raggedright\arraybackslash}p{5cm}}
\newcolumntype{Y}{>{\raggedright\arraybackslash}p{4cm}}
\newcolumntype{X}{>{\columncolor{VA}\ttfamily\small}p{9cm}}
\newcommand{\pquotes}[1]{\textcolor{gray}{\textit{#1}}}

\def\eg{\emph{e.g., }}

\AtBeginDocument{%
  }

\begin{document}

\title[From Our Lab to Their Homes: Learnings from Longitudinal Field Research with Older Adults]{From Our Lab to Their Homes: \\Learnings from Longitudinal Field Research with Older Adults}

\author{Amama Mahmood}
\email{amama.mahmood@jhu.edu}

\affiliation{%
  \institution{The Johns Hopkins University}
  \streetaddress{3400 N. Charles St}
  \city{Baltimore}
  \state{Maryland}
  \country{USA}
  \postcode{21218}
}

\author{Chien-Ming Huang}
\email{chienming.huang@jhu.edu}
\affiliation{%
  \institution{The Johns Hopkins University}
  \streetaddress{3400 N. Charles St}
  \city{Baltimore}
  \state{Maryland}
  \country{USA}
  \postcode{21218}
}
\renewcommand{\shortauthors}{Mahmood \& Huang}

\begin{abstract}
Conducting research with older adults in their home environments presents unique opportunities and challenges that differ significantly from traditional lab-based studies. 
In this paper, we share our experiences from year-long research activities aiming to design and evaluate conversational voice assistants for older adults through longitudinal deployment, interviews, co-design workshops, and evaluation studies. 
We discuss the benefits of bringing the lab to their home, including producing realistic and contextual interactions, creating stronger researcher-participant bonds, and enabling participant growth with the research over time. We also detail the difficulties encountered in various aspects of the research process, including recruitment, scheduling, logistics, following study protocols, and study closure. 
These learnings highlight the complex, yet rewarding, nature of longitudinal home-based research with older adults, offering lessons for future studies aiming to achieve real-world applicability.
\end{abstract}


\begin{teaserfigure}
  \includegraphics[width=\textwidth]{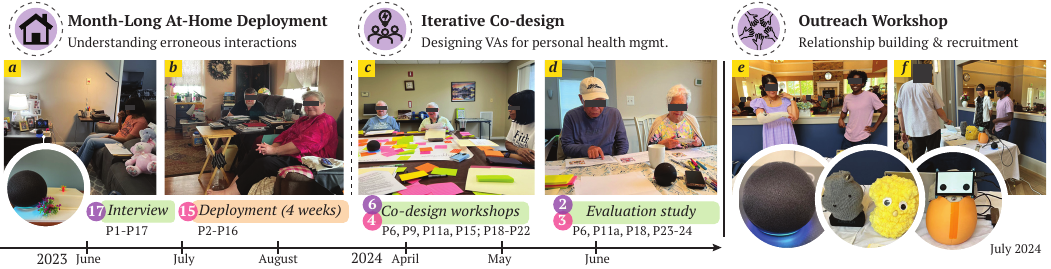}    
     \caption{Timeline of research activities for projects 1 (a, b) and 2 (c, d), and outreach workshop (e, f). Purple dots represent the number of new participants, and pink dots represent the number of returning participants. Participant codes are consistent across all sessions; for example, P15 in interviews, the month-long deployment, and the co-design workshop is the same individual. P10, P11, P12, P16, and P22 are couples.}
    \Description{The figure shows the timeline of research activities for two studies and outreach activities. 
    The timeline (labelled at the bottom) goes from June 2023 till June 2024. The first study is month-long in-home deployment to understand erroneous interactions between voice assistants and older adults. The second study is iterative co-design for designing VAs for health self-management. The timeline from left to right goes as: 
    1. June 2023 are interviews for second project with P1-P17 and a tag indicating that all participants are new. 
    2. The deployment spans from June till August 2023, 4 weeks each for participants P2-P16. A tag indicates that all participants were subset of the interview group. The picture a and b shows the deployed device with participants in their house. 
    3. The co-design workshops for project two starts in April 2024 and ends in May 2024 with 4 participants from the previous activities (P6, P9, P11a, and P15) and six new participants (P18 to P22).
    4. The evaluation study occurred in June 2024 with 3 participants from previous activities (P6, P11a, P18) and two new participants (P23, P24). 
    On the right side of the figure, the outreach workshop is shown, The research team can be seen interacting with participants in the figures. There are multiple robots and a smart speaker on the table to demonstrate to the older adults.}
  \label{fig:teaser}
\end{teaserfigure}

\maketitle

\renewcommand{\thefootnote}{}
\footnotetext{This is the author’s version of the work. It is posted here for your personal use. Not for redistribution.}

\renewcommand{\thefootnote}{\arabic{footnote}}

\section{Introduction}

To design technology for enhanced quality of life and to support aging in place, it is important to consider three strands of research design: involving older adults, conducting research at their homes, and for longer periods of time.
First, the research agenda for assistive technologies should be shaped by older adults through deeper researcher-community partnerships, which improve planning, execution, and outcomes \cite{vines2015age}. Building relationships with older adults is crucial and can be facilitated through interactive activities and workshops in their personal spaces \cite{binda2018recruiting}. Second, conducting research in `artificial' lab environments limits the observation of contextualized use \cite{franz2019usability}. Whereas, conducting research in the comfort of older adults' homes helps identify realistic usability issues and accessibility barriers, ensuring technology better fits their physical, cognitive and sensory needs \cite{ostrowski2021long, ostrowski2022mixed}. Third, longitudinal research captures long-term lived experiences, providing insights into gradual processes such as aging and learning \cite{ostrowski2021long}.

Despite the importance of longitudinal in-home research with aging adults, such studies remain uncommon. A recent review identified only 21 longitudinal studies in older adults' homes with voice assistants (VAs), with many usability and evaluation issues still unaddressed \cite{arnold2024does}. 
Plethora of challenges including recruitment difficulties, participant engagement, and the lack of long-term engagement methodologies \cite{binda2018recruiting, ostrowski2021long}, hinders longitudinal in-home research. This paper expands on the benefits and challenges of in-home longitudinal research with older adults, drawing from our experiences over the past year with older adults to design and evaluate advanced conversational voice assistants through interviews, co-design workshops, and evaluation studies. We expand on prior work that focused on recruitment and initial engagement \cite{binda2018recruiting} by \textit{exploring participant retention across multiple research projects highlighting the benefits of long-term participation. 
We juxtapose these benefits with the challenges of longitudinal in-home research and present suggestions for addressing these challenges. We argue for long-term involvement of older adults as co-creators of assistive technology to support aging in place.}

Our discussion below draws from our experience with two in-home projects involving older adults, focusing on designing and evaluating voice assistants (Figure \ref{fig:teaser}). The first was a month-long study ($N=15$), aimed at understanding interaction breakdowns with VAs.
This study involved deploying Alexa with older adults for 4 weeks and  adding an Alexa skill with ChatGPT for multi-turn conversations for the last 2 weeks. At the end of the study, we explored the integration of ChatGPT into Alexa for self-diagnosis in a structured session where participants received printed interaction instructions and symptom descriptions. The second project involved co-designing a personal health assistant with older adults through various steps: 1) an interview study ($N=17$) to identify challenges in healthcare self-management, 2) three co-design workshops ($N=10$) to develop a large language model (LLM) powered prototype VA, and 3) an evaluation study ($N=5$) to assess the usability and effectiveness of the designed VA.

\section{Benefits of Bringing Lab to Home}
Over a year, we observed older adults' interactions within and outside the study context. We noted several benefits of conducting longitudinal in-home research across projects.

\subsection{Realism and Attitude Shift}
\label{sec:benefits-realism}

Bringing the lab to the home environment provides a context that is more authentic and reflective of daily life. Interactions with voice assistants are organic; 
during our month-long deployment, we observed how participants seamlessly integrated Alexa into their daily routines. For example, 
a couple used Alexa to answer an open question, where P16a verified Alexa's answer to P16b's question (Table \ref{tab:conversations}). 
\begin{table}[h]
\centering
\caption{Example participant conversation}
\begin{tabular}{M}
\textit{\textbf{P16b:}}  Did the actress who played the voice of Howard Wolowitz on third rock from the sun really die?  \\
\cellcolor{VA} \textit{\textbf{VA:}}  From\dots: But sadly the actress behind that memorable voice has passed away.  \\
\textit{\textbf{P16a:}} That is true. She did pass away. \\

\end{tabular}
\label{tab:conversations}
\end{table}

The home setting also allowed us to obtain unfiltered, realistic feedback. 
The interaction logs from our deployment study captured both negative and affectionate remarks directed at Alexa that participants might be reluctant to express in a lab setting. Negative feedback often included sarcastic comments like \pquotes{``hell of a job''} or critical remarks such as \pquotes{``Great answers for questions I didn't ask''} and \pquotes{``You're a circular idiot.''} Conversely, participants also expressed affection, with statements such as \pquotes{``I really like you.''} and \pquotes{``I love you.''}

Additionally, we observed a noticeable shift in the participants' attitudes due to the comfort of being in their personal space; older adults approached tasks differently at home compared to a lab setting---even in the absence of an experimenter in lab or with the presence of an experimenter at home. For example, during a structured self-diagnosis task using an LLM-powered VA conducted at the commencement of the month-long study, older adults focused more on their own health issues and general medical advice rather than strictly following the prescribed symptoms given by the experimenter.
This shift suggests that being in a familiar environment may make them feel safe and encourage more open and personal conversations even about sensitive topics such as health.

\subsection{Stronger Researcher-Participant Relationship}
\label{sec:stronger-bond}
Inviting researchers into their homes indicates a stronger bond between participants and researchers compared to lab settings, where interactions can be more impersonal---experimenter and participant may not even know each other's name. On the contrary, the home-based setting fosters a more personal and engaging relationship.

\subsubsection{Extended and repeated interactions} 
One significant aspect of this relationship is the extended interaction time. In both of our studies, 
we spent considerable time with participants ($\approx$3 hours each). 
This extended interaction allowed for deeper connections and a better understanding of participants' needs and behaviors.

Another crucial aspect is the repeated interactions throughout a study and sometimes across studies. This continuous contact not only strengthens the relationship but also builds trust that fosters open communication. During our month-long deployment study, participants frequently shared their experiences with the system by sending unprompted text messages to the researchers;  
for example, one participant expressed gratitude via texting, \pquotes{``I am thanking Alexa. Now if she could just take a shower and wash my hair for me.''} They reached out to discuss both issues they were facing and new features they were exploring. 

\subsubsection{Beyond the study: continued support and personal connection}
Researcher-participant interactions were not confined to the study itself. Participants often asked for help with other technology related issues.
Even after the study ended, some participants continued to contact the experimenter for technical support, leading to jokes in the research group about the experimenter becoming their ``tech support for life.'' While this ongoing communication may be desirable for furthering research with participants, measures should be taken to protect researchers. For instance, using designated work phones and emails for communication can help avoid violating the personal space of researchers.

Similarly, communication is not only limited to assistance with technology. Participants also shared personal stories and experiences. They inquired about the researcher's future plans and even suggested hosting a graduation party for them. Such exchanges highlighted a genuine interest in mutual understanding and support. Prior work has also shown that mutual trust and informal conversations are important for involving older adults in research process \cite{binda2018recruiting}; particularly, interactive games and other fun activities between participants and researchers were found effective for gathering data informally.  Establishing rapport can help retain participants and maintain long-term relationships that are beneficial for co-designing technologies with them \cite{ostrowski2021long}.

\subsubsection{Long-term engagement: participants contributing to recruitment}
When approached for recruitment of later projects, participants often shared positive feedback about their prior experiences and how they are enjoying their VA almost as a way to thank the researchers for introducing them to this new technology and showcasing their willingness to be further involved in research activities. For instance, 
a participant (P15) enthusiastically mentioned that she now has two more smart speakers since after the month-long deployment, \pquotes{``I now have two more Alexas. One in my office and one in my bedroom. She reminds me when to take my medications, when to leave for meetings, and what's on my calendar for the day.''} Participants also actively assisted in recruitment and logistics for studies---helping to schedule and arrange community meeting spaces for workshops, calling potential participants, and promoting the research to family, friends and acquaintances.

\subsection{Growing Together: Maintaining Research Momentum} 
\label{sec:growing-together}
The stronger bond cultivated through longitudinal home-based research provides opportunities for the inclusion of older participants in a long-term research agenda involving multi-step and diverse projects.



\subsubsection{Evolving mental models: understanding capabilities and purposes}
As research progresses, participants' continued involvement allows them to grow alongside. Researchers gain insights from participants, while participants not only learn to use new systems but also evolve their understanding of the systems' purposes with each interaction. They gradually contextualize how these systems can further be integrated into their daily lives, benefiting from a step-wise learning process rather than a sudden introduction to novel technologies. Our experience shows that older adults, often unfamiliar with technologies such as smartphones and personal VAs, face inherent challenges in understanding the potential of a voice assistant if they do not fully grasp its capabilities. Thus, gradual learning through various research activities fills this gap by helping them become comfortable with the technology, allowing them to form and evolve their mental model of these technologies over time and build capacity to participate in the process as co-designers \cite{ostrowski2021long}.


\subsubsection{From insight to innovation: envisioning future capabilities}
Involving older adults throughout the research pipeline or across various projects not only expands their knowledge of these technologies' capabilities but also enables them to envision additional applications and functionalities.  This trend was particularly evident during co-design workshops with a high-fidelity prototype of an LLM-powered VA for personal healthcare management. We had a mix of participants: those with prior experience with Alexa through our month-long deployment and those with no prior experience. 
The participants who had been part of our earlier studies could extrapolate the VA's capabilities and suggest actionable features due to their long-term interactions with VAs. 
For example, one participant (P11a), who previously participated in the longitudinal study, suggested that the VA could adjust medication reminders based on personal calendars in case there are disruptions in user's routine, drawing from his experience with Alexa reading personal calendar events. 
During the co-design workshop, a participant without prior study experience asked if the VA could provide automatic notifications about high pollen levels that may necessitate taking allergy medications. More experienced participants (P9 and P11a) were able to present an actionable extrapolation that since Alexa can already give weather updates and pollen counts on request, it would be possible for it to monitor the pollen count and notify the user to take allergy medication if needed. 
Thus, the benefits of retaining same participants across research projects are similar to experience-based co-design \cite{harrington2018designing} where participants first try out the technology for certain amount of time before coming into the co-design process \cite{harrington2018designing}. However, continued involvement can accelerate the research process by eliminating the need for new participants to familiarize themselves with the technology.

\subsubsection{Leveraging experience to bridge functionality gaps}
On many occasions, experienced participants relied on their interactions with Alexa since the longitudinal study (project 1) to suggest solutions during the co-design workshop (project 2). For instance, P6 suggested that the personal health assistant should remind users about prescription refills, similar to how Alexa reminds about over-the-counter medications, \pquotes{``I have found Alexa to be very useful because of the Amazon shopping connection. What she does is remind me when I need a refill on a medication and I order it from Amazon. She tells me the number of pills in the bottle and the cost, and all I say is 'yes, Alexa.' It's done and it comes. That's very handy.''} This suggestion highlights how participants can draw parallels from their use of VAs for one task to another.  

Ongoing learning and growing together enable participants to articulate what works and what does not work for them and suggest continual improvements. For instance, one participant (P15) identified a breakdown in Alexa's calendar event notification during the co-design workshop, emphasizing the need for Alexa to provide more detailed reminders, specifically sensitive reminders \eg doctor's appointment: \pquotes{``Alexa must read my calendar. It must read my calendar because a lot of times it'll tell me like when I get up in the morning, I say Good morning, my good morning routine is that it tells me what's on my calendar\dots [Alexa says] You have a meeting at [XYZ] at 1 o'clock tomorrow. I'm like okay [with] who. I may have to find out who but thanks for reminding me.''} 

\subsubsection{Experience to expectation: transferable learning across technologies}
We observed the benefits of long-term participation across different technologies. For instance, in a different project, when discussing the use of assistive robots 
for various applications, participants of the VA studies pointed out how an embodied robot should differ from a VA and offer more advanced features to fully exploit its embodiment; otherwise, it would be redundant and undesired. 
They could visualize what embodiment could be better for. Thus, long-term learning is not confined to a specific technology but is transferable, enhancing participants' overall awareness of various assistive technologies and encouraging them to think critically about their needs and expectations. Their expectations from a certain type of system become more grounded in their understanding of various technologies. 

\subsubsection{Differences in prior experience: personal use vs. research studies}
One may ask why not simply include more people with prior experience with VAs instead of retaining participants from previous studies? Prior work shows that experienced technology users are far better prepared to engage in participatory design than novice users \cite{pater2017addressing, pollack2016closing}.
However, we observed that participants with personal  prior experience using even multiple Amazon Echo devices did not know many of its capabilities before our study and were less inclined to try new things, as they might be set in their ways of usage. In contrast, participants new to the technology were eager to learn and experiment. One participant (P15), during the co-design workshop, mentioned, \pquotes{``For me, it's kind of like the first time I started using her. It's because you kept feeding me things and I kept figuring out what. I couldn't figure out what she could do. But you kept saying try this this way and I was like oh and that made me go I can do this let me try this let me try that and you know when you kept feeding me, I kept figuring out things, you know. Oh wow, she can do that.''}
Other participants continued to find their own uses for the technology during our month-long study. For instance, P6 enjoyed getting sports updates from Alexa, and P2 tested Alexa's capabilities for information retrieval. This highlights that recruiting people with prior personal VA experience is different from recruiting those who have participated in similar studies before.

How technology is introduced to people affects adoption. 
During early adoption stages, exploration can be hindered by users' low familiarity with technology and their inability to learn independently. However, participants introduced to VA technology through our study received extensive demonstrations and were encouraged to explore. 
Such experimenter interactions can boost curiosity and exploration, increasing the chances of continued use for various purposes. Therefore, retaining and cultivating participants with no prior personal experience through continued research has its benefits.

To summarize, growing and learning together helps propagate research ideas faster and more organically, preventing participants from always playing catch-up with researchers. It mitigates the challenge of adopting novel technologies and supports continued research and co-design efforts with the target population.
However, it is still essential to include new participants at different stages of the same or different research to ensure diverse perspectives and a representative sample to ensure that we embrace diversity and avoid treating aging adults as a homogeneous group as sometimes done in the HCI community \cite{vines2015age}.

\section{Challenges of Conducting In-Home Research}
While conducting longitudinal in-home research offers benefits, we also encountered numerous challenges. Below, we describe these challenges, the actions we took to address them, and provide suggestions for future research.
\subsection{Recruiting Older Adults for In-Home Studies}
\label{sec:recruitment}
Recruiting older adults in general is challenging, and the difficulties are manifold for in-home research. Our experience has shown that traditional methods, such as distributing flyers and utilizing mailing lists, are insufficient. Below, we discuss different aspects of recruitment: 

\subsubsection{Recruitment channel} 
To recruit community-dwelling older adults, 
we initially used our connections at the nursing school and posted flyers in the vicinity of the university campus. Over time, we formed relationships and leveraged snowball sampling and word-of-mouth referrals to enhance our recruitment efforts. While the initial contact is relatively fast for individuals, the recruitment numbers are low. 

To increase recruitment numbers, we established connections with community centers, though this approach has its own difficulties due to the high older adults-to-caregiver ratio. We reached out to approximately ten community centers and received responses from only three, ultimately establishing and maintaining a partnership with just one over the course of a year. The partnership has been meaningful---the administration of the independent and assisted living community center has been very forthcoming and helpful in recruitment and scheduling for the studies. The lead researcher was also invited to speak to the residents of the independent living facility at their monthly town hall meeting, where she introduced her research and initiated the recruitment process.

However, in their assisted living facility, the administration often acts as an intermediary, ensuring the safety and well-being of residents. This involvement is crucial but also adds a layer of complexity to the process---staff availability can be limited due to high resident-to-caregiver ratios, making it difficult to coordinate schedules and maintain consistent communication.
While it is difficult establishing a lasting collaboration, but once it is set up, the subsequent recruitment efforts become easier. 

\subsubsection{Tunnel vision: homogeneous participant pool}
Maintaining relationships with participants aids in recruitment, but there is a concern that this approach can lead to tunnel vision, as it may result in a homogeneous participant pool with similar contexts due to their social circles leading to a narrow range of perspectives and experiences.
To mitigate this tunnel vision effect, we have recently increased our recruitment drives and outreach efforts to attract a more diverse group of older adults. We conducted an interactive exhibition/workshop  at a local community center (see Figure. \ref{fig:teaser}) 
featuring technology such as robots and smart speakers. The residents were excited and wanted to be part of our research. We exchanged contact details with interested individuals to keep them informed about upcoming projects. 

We need to focus more on outreach activities specifically to recruit diverse participants. For instance, general recruitment materials and practices may be suboptimal for recruiting immigrants, as they may be shy and reluctant due to language and cultural barriers. We faced this issue while recruiting immigrant participants for our co-design workshops, where we sought their unique perspectives on VAs for health self-management. We had to rely on convenience sampling by reaching out to people the research team was familiar with, such as family and friends. 
To facilitate communication, a member of our research team, fluent in Mandarin Chinese, communicated with mandarin-speaking participants (P22 couple) throughout the workshop. Seeking diverse perspectives is necessary for inclusive research, but it often incurs higher costs due to the need for translators and facilitators. To enable inclusive research, we suggest: 1) creating culturally sensitive recruitment materials and outreach activities, 2) collaborating with local specialized organizations and community leaders, 
3) engaging translators or bilingual researchers as needed, and
4) using personal invitations and follow-ups to build rapport and make participants feel valued and comfortable.

\subsection{Logistics}
\label{sec:logistics}
We faced many logistical challenges  in time, effort, and resources while conducting in-home research with older adults:

\subsubsection{Resources}
Human resource (\eg researchers' time) and transportation costs associated with conducting in-home studies are substantial.
For example, during our month-long study involving 15 deployments, the transportation costs amounted to approximately \$500. These deployments were completed in four waves, and distance-wise, only three of the deployments were within walking distance.
The researcher typically spent around three hours with each participant, divided into 90-minute sessions at the start and end of the deployment. 
To mitigate the transportation cost, we should consider developing packageable technologies that can be mailed to participants. For instance, we could mail smart speakers with easy-to-setup instructions and the ability to have a phone call or video chat with research team.

\subsubsection{Setting up devices}

Unlike in-lab studies, experimenters have limited control over the situation in participants' homes making the integration of devices into their ecosystem challenging. The researchers should always prepare themselves for unexpected \cite{binda2018recruiting}. 
For instance, one common issue faced by us is dealing with internet connectivity. A significant amount of time, sometimes up to 30 minutes, is often spent figuring out the participants' internet name and password. Participants are usually unaware of their existing systems, requiring experimenters to work closely with participants to resolve these issues.
Providing clear instructions and requirements beforehand can mitigate these challenges but might discourage participation due to the conscious effort required on their part. 
During scheduling calls, participants often asked if they had to do anything specific, expressing concern about potential inconvenience. 

\begin{table*}[t]
\centering
\caption{Summary of benefits, challenges and suggestions for longitudinal in-home research with older adults. }
\label{tab:summary}
\begin{tabular}{Y S L}

\textbf{Benefits}              & \textbf{Challenges}                          & \textbf{Suggestions} \\ 
\hline
\textbf{\ref{sec:benefits-realism} Realism} & \textbf{\ref{sec:protocol} Organic interaction vs. protocol} & \\
Contextual and organic interactions provide realistic insights & Participants may not follow the protocol exactly in the comfort of their homes & Researchers should balance encouraging organic interaction and maintaining protocol \\ 
\hline
\textbf{\ref{sec:stronger-bond} Stronger researcher- participant bond} & \textbf{\ref{sec:logistics} Logistics} & \\
Repeated in-home interactions results in strong bond
& Human resources and transportation costs; hard setting up devices & Create packageable and shippable technologies with easy setup guides  \\ 
\lighthline
Stronger research- participant bond creates continued mutual support &  Invades researchers' personal space and takes up their time & Designated work phone/email to draw boundaries yet being accessible   \\ 
\hline
\textbf{\ref{sec:growing-together} Growing together} & \textbf{\ref{sec:recruitment} Recruitment} & \\
Participation across studies can transition older adults into true co-designers and co-creators &Hinders diverse perspective and restricts co-designed technology to experienced participants & 
Include a mix of experienced and new participants to leverage the insights of seasoned users and the fresh perspectives of newcomers \\ 
\lighthline
Established community connections 
aid in recruitment through snowball sampling & Creates tunnel vision, resulting in a homogeneous participant pool & Conduct community outreach activities such as interactive expos and exhibitions featuring research tools (\eg robots) \\
\lighthline
Convenience sampling through personal connections to recruit special groups & Recruitment numbers are very low and may have personal biases. Generic recruitment materials do not fit all & Tailor recruitment materials to specific population and  collaborate with local organizations and specialized centers for targeted outreach \\ 
\hline
& \textbf{\ref{sec:attachement} Attachment with Agents} & \\
Participants' bond with agents is beneficial for the adoption and continued use 
& Participants may feel an emotional toll if agents are taken away at the end of research activities & Develop inexpensive systems, offer affordable alternatives, or incorporate debrief sessions to help users transition \\

\hline
\end{tabular}

\end{table*}

\subsection{Organic Interaction vs. Protocol: A Double-Edged Sword}
\label{sec:protocol}
People at home do not necessarily follow protocols as strictly as in controlled lab-based studies. We observed that older adults at home approached a controlled self-diagnosis task with an LLM-powered VA very differently compared to young adults in a lab setting. Instead of sticking to the given symptoms, they focused more on their own health issues while asking the VA self-diagnosis questions and seeking general medical advice. 
During the session, only 2 out of 12 older adult participants adhered strictly to the script. Four started with the script but deviated to personal questions, while the remaining six asked questions related to their own health concerns from the start. In contrast, all 20 younger participants in a similar lab study followed the script exactly. Additionally, we observed that participants asked more personal questions when at home, either alone or with a spouse, compared to when in a group setting during the co-design workshops.
The older adults' tendency to ask questions based on their personal health concerns led to them to go off-script. For example, P7 asked about muscle ache and soreness, P9 had questions related to yoga and trouble sleeping, P12b asked about medication interactions and P13 discussed 
and leg cramps at night. 
Frequent deviations from the protocol highlights the challenge of maintaining controlled conditions in in-home studies. 
Extra precautions may be needed to ensure older adults follow the protocol at home, as lab visits would be even harder to arrange.

On the flip side, as discussed above, 
divergence from protocol can also be informative as it provides insights into authentic use in a real-world context. We observed that participants found the information provided by the VA to be helpful and expressed their appreciation directly to the VA. For instance, P14 said, \pquotes{``Thank you, Alexa, that was so helpful,''}  indicating sincere interest in the information rather than in a contrived scenario. Such genuine engagement led to more meaningful conversations, providing us deeper insights into how they might use the VA naturally, as opposed to if they had strictly followed the protocol. 
This presents as a double-edged sword: on one hand, organic interactions provide realistic insights; on the other hand, deviations from the protocol can compromise the consistency of the study. 
Thus, we should balance encouraging organic interactions and maintaining adherence to protocol.

\subsection{Attachment with Agents}
\label{sec:attachement}
Older adults may develop attachments to the agents or assistants they interact with. After the month-long study, participants were given the option to either keep the smart speaker or receive compensation equivalent to the market price of an Amazon Echo Dot. All participants except three chose to keep the smart speaker. 

Attachment to VAs is evidenced by social niceties such as expressing gratitude and greetings, a phenomenon also noted in prior research \cite{pradhan2019phantom}. 
During our month-long study, interaction data showed that participants frequently expressed gratitude,
greeted Alexa, 
and laughed at its responses.
Participants often attributed human characteristics to Alexa because of the way it responded to social interactions. For example, P14 mentioned: 
\pquotes{``When I say, `thank you Alexa,' Alexa will say, `oh, you are welcome, I'm here to serve you.' I don't know why I like to hear that as if it's a real person or something.'' }
Similarly, previous studies have shown that older adults are more likely to anthropomorphize the agent, using polite language to express gratitude, while younger adults tend to view it more as a practical tool, focusing on its functionality \cite{oh2020differences, chung2019elderly}. Similar to prior work  \cite{chung2019elderly}, participants' perceptions of Alexa shifted from viewing it as a tool to seeing it as a friend and companion over time as P15 stated, \pquotes{``I think in the beginning I thought of it as a thing and then when I started saying nice stuff and I started saying nice stuff, then I just think of it as Alexa. So it's like another, you know, I didn't have anybody to get up in the morning and say `Good morning' to, so I can get up and say `Good morning Alexa' and she goes: `GOOD MORNING L. Today on your calendar.' So, I sit there and I listen to her, so yeah it gives her a personality because in the beginning I kind of felt like it is intruding but now it's like `Hey girl'.''}
This tendency makes older adults more susceptible to forming emotional attachments  
that may not only lead to heightened expectations but also raise ethical concerns, including issues of deception, infantilization, and privacy \cite{sharkey2012granny, wachsmuth2018robots}. Given that older adults are a more vulnerable population, it is essential to design study protocols with these potential consequences in mind. 


As researchers, we should consider the emotional toll it may have on participants if the agent is taken away after they have grown fond of it, especially in longitudinal studies. We may consider using or creating inexpensive agents to be able to leave it with them if they desire. However, since it is not always possible to provide participants with the devices permanently, other ways should be considered to reduce the emotional impact. For instance, offering similar affordable alternatives, providing extended access when possible after conclusion of the study, 
or incorporating debriefing sessions to help participants transition away from the use of the technology. Additionally, fostering ongoing communication and support can help mitigate the sense of loss and maintain a positive relationship with the participants.

\section{Concluding Remarks}

In this paper, we juxtapose the benefits and challenges associated with much-needed in-home longitudinal research with older adults (summarized in Table \ref{tab:summary}). There are clear benefits in establishing stronger bonds to encourage long-term participation, allowing older adults to grow with the research and become true co-designers and co-creators. We highlight numerous logistical challenges that make longitudinal field research time- and resource-consuming, and touch on the possible emotional toll of taking agents away after research is over. We describe the actions we took to address these issues and propose possible strategies. Additionally, we point out that some benefits and challenges are tangled and present as double-edged swords. For instance, building rapport through personal connections can enhance participation but may compromise researchers' time. Similarly, organic interactions provide valuable insights, but deviations from the protocol can cause inconsistencies. Hence, it is a balancing act that researchers should be mindful of while conducting in-home longitudinal studies with aging population.

    
\section*{Acknowledgements}
This work was supported by the National Science Foundation award \#1840088 and the Malone Center for Engineering in Healthcare.

\section*{CRediT author Statement}
\textbf{AM}: Conceptualization, Methodology, Software, Validation, Formal analysis, Investigation, Data curation, Writing - Original draft, Writing - Review \& editing, Visualization, Project Administration. 
\textbf{CMH}: Conceptualization, Methodology, Resources, Writing - Original draft, Writing - Review \& editing, Visualization, Supervision, Funding acquisition. 

\bibliographystyle{ACM-Reference-Format}
\bibliography{references}
\end{document}